\documentclass[preprint,showpacs,showkeys,preprintnumbers,amsmath,amssymb]{revtex4}

\usepackage{graphicx}% Include figure files
\usepackage{dcolumn}% Align table columns on decimal point
\usepackage{bm}% bold math

\def\half{{1\over 2}}

\def\u#1{{\bf #1}}

\def\rdot{\dot r}

\begin{document}

%\preprint{APS/123-QED}

\title{A New Superintegrable Hamiltonian}

\author{P.E. Verrier}
\email{pverrier@ast.cam.ac.uk}
\author{N.W. Evans}
\email{nwe@ast.cam.ac.uk}
\affiliation{%
Institute of Astronomy, Madingley Rd, University of Cambridge, CB3 0HA, UK
}

\date{\today}

%%%%%%%%%%%%%%%%%%%%%%%%%%%%%%%%%%%%%%%%%%%%%%%%%%%%%%%%%%%%%%%%%%%%%%
%%%%%%%%%%%%%%%%%%%%%%%%%%%%%%%SECTION%%%%%%%%%%%%%%%%%%%%%%%%%%%%%%%%
%%%%%%%%%%%%%%%%%%%%%%%%0%%%%%%%%%%%%%%%%%%%%%%%%%%%%%%%%%%%%%%%%%%%%%%

\begin{abstract}
  We identify a new superintegrable Hamiltonian in 3 degrees of
  freedom, obtained as a reduction of pure Keplerian motion in 6
  dimensions.  The new Hamiltonian is a generalization of the
  Keplerian one, and has the familiar $1/r$ potential with three
  barrier terms preventing the particle crossing the principal planes.
  In 3 degrees of freedom, there are 5 functionally independent
  integrals of motion, and all bound, classical trajectories are
  closed and strictly periodic.  The generalisation of the
  Laplace-Runge-Lenz vector is identified and shown to provide
  functionally independent isolating integrals. They are quartic in
  the momenta and do not arise from separability of the
  Hamilton-Jacobi equation. A formulation of the system in
  action-angle variables is presented.
\end{abstract}

\pacs{Valid PACS appear here}
\keywords{Suggested keywords}

\maketitle

%%%%%%%%%%%%%%%%%%%%%%%%%%%%%%%%%%%%%%%%%%%%%%%%%%%%%%%%%%%%%%%%%%%%%%
%%%%%%%%%%%%%%%%%%%%%%%%%%%%%%%SECTION%%%%%%%%%%%%%%%%%%%%%%%%%%%%%%%%
%%%%%%%%%%%%%%%%%%%%%%%%%%%%%%%%%%%%%%%%%%%%%%%%%%%%%%%%%%%%%%%%%%%%%%

\section{\label{sec:intro}Introduction}

The Kepler problem is well known to be superintegrable -- that is, it
has five functionally independent integrals of motion. They are the
energy and the components of angular momentum and the
Laplace-Runge-Lenz vectors, obtainable by separating the
Hamilton-Jacobi equation in spherical polar and rotational parabolic
coordinates~\citep{La69}. For Hamiltonians with three degrees of
freedom, the existence of five integrals of motion implies that every
bound trajectory is closed.

Sommerfeld and Born, in the days of the old quantum theory, appear to
have been the first to realize that if a potential is separable in
more than one coordinate system, it possesses additional isolating
functionally independent integrals~\citep{So23,Bo27}.  The first
systematic inquiry into this problem was begun by Winternitz,
Smorodinsky and co-workers, who found every potential in two degrees
of freedom for which the Hamilton-Jacobi equation is separable in more
than one way~\citep{Fr65}. Subsequently, they extended this work to
three degrees of freedom by finding every potential separable in
spherical polars and at least one additional coordinate
system~\citep{Ma67}. Evans~\citep{Ev90} then completed this work by
investigating all the remaining possibilities. A useful introduction
to the subject of superintegrability, as well as summary of recent
work, is given in the conference proceedings of Tempesta et
al.~\citep{Te05}.

If the Hamilton-Jacobi equation separates, then the corresponding
integral of motion is necessarily linear or quadratic in the canonical
momenta. Consequently, all the superintegrable systems listed
in~\citep{Fr65,Ma67,Ev90} have integrals that are quadratic in the
momenta. As an example, let us consider the Keplerian Hamiltonian
\begin{equation}
H = \half|\u{\rdot}|^2  -\frac{k}{r},
\label{eq:Vkep}
\end{equation}
where $k$ is a real positive constants. Letting $\u{r}$ denote the
position vector, then the integrals are the energy $E$, the components
of the angular monetum vector $\u{L} = \u{r} \times \u{\rdot}$ and the
Laplace-Runge-Lenz vector 
\begin{equation}
\u{A} = \u{\rdot} \times \u{L} - \frac{k}{r}\u{r},
\end{equation}
all of which are at most quadratic in the velocities. 

Superintegrable systems with higher-order integrals are known
~\cite{Gr02}, although they are extremely scarce. Examples include the
anisotropic harmonic with rational frequency ratio $\ell : m: n$ where
$\ell + m + n \ge 5$, i.e.,
\begin{equation}
H = \half |\u{\rdot}|^2 + \ell^2 x^2 + m^2 y^2 + n^2 z^2.
\label{eq:anosc}
\end{equation}
The potential separates in rectangular cartesians and possesses two
commuting quadratic integrals. There are two additional integrals
which may be taken as polynomials of degree $\ell + m - 1$ and $\ell +
n -1$~\citep{Ja40}. The Calogero potential in a harmonic well
\begin{equation}
H = \half |\u{\rdot}|^2 + k(x^2 + y^2 + z^2) +{k_1\over (x-y)^2}
+ {k_1\over (y-z)^2} + {k_1\over(z-x)^2},
\end{equation}
is known to be super-integrable and possesses an integral of the
motion that is cubic in the velocities \citep{Ad77,Wo83}.

In this paper, we introduce a new superintegrable Hamiltonian, namely
\begin{equation}
H = \half|\u{\rdot}|^2  -\frac{k}{r} + \frac{k_1}{x^2} + \frac{k_2}{y^2} 
+ \frac{k_3}{z^2}.
\label{eq:Vgenkep}
\end{equation}
This is recognized as a generalization of the familiar Keplerian
Hamiltonian.  The constants $k_1, k_2$ and $k_3$ are taken as
positive, so the Hamiltonian is perfectly physical and motion is
confined to, say, the octant $x >0, y>0 $ and $z >0$. The terms
involving the $k_i$ correspond to repulsive barriers preventing the
orbit crossing any of the principal planes.  In Section~\ref{sec:evi},
we present numerical integrations showing that the orbits give closed
curves in all cases. This motivates a search for the isolating
integrals, one of which is found to be quartic in the momenta in
Section~\ref{sec:int}.  Finally, an action angle formalism is given in
Section~\ref{sec:acang} and the relation to the Kepler problem
discussed.

%%%%%%%%%%%%%%%%%%%%%%%%%%%%%%%%%%%%%%%%%%%%%%%%%%%%%%%%%%%%%%%%%%%%%%
%%%%%%%%%%%%%%%%%%%%%%%%%%%%%%%SECTION%%%%%%%%%%%%%%%%%%%%%%%%%%%%%%%%
%%%%%%%%%%%%%%%%%%%%%%%%%%%%%%%%%%%%%%%%%%%%%%%%%%%%%%%%%%%%%%%%%%%%%%

\section{Evidence of Superintegrability}
\label{sec:evi}

\subsection{Analytic Proof}

Let us recall that Keplerian motion in $N$ degrees of freedom always
possesses $2N-1$ functionally independent integrals of
motion~\citep{Mo}. Specialising to 6 degrees of freedom, we have the
Hamiltonian
\begin{equation}
H = \half |\u{p}|^2 - {k\over |\u{s}|}
\label{eq:six}
\end{equation}
where $s$ has Cartesian coordinates $(s_1,s_2,s_3,s_4,s_5,s_6)$. Now,
let us introduce coordinates $(x,y,z,\theta_x, \theta_y,\theta_z)$
according to
\begin{eqnarray}
s_1 &=& x \cos\theta_x, \qquad s_2 = x \sin \theta_x \nonumber\\
s_3 &=& y \cos\theta_y, \qquad s_4 = y \sin \theta_y \nonumber\\
s_5 &=& z \cos\theta_z, \qquad s_6 = z \sin \theta_z \nonumber
\end{eqnarray}
The Hamiltonian becomes
\begin{equation}
H = \half( p_1^2 + p_2^2 + p_3^2 + {p_{\theta_1}^2\over x^2}
+ {p_{\theta_2}^2\over y^2} + {p_{\theta_3}^2\over z^2})
 - {k\over (x^2 + y^2 + z^2)^\half}\nonumber
\end{equation}
The coordinates  $(\theta_1, \theta_2, \theta_3$) are ignorable, so
we obtain a new Hamiltonian
\begin{equation}
H = \half (p_1^2 + p_2^2 + p_3^2) - {k\over r} + {k_1 \over x^2} +
{k_2\over y^2} + {k_3 \over z^2}
\label{eq:three}
\end{equation}
where $k_1, k_2$ and $k_3$ are the constant values of the momenta
conjugate to the ignorable coordinates. In the original 6 degrees of
freedom Hamiltonian~(\ref{eq:six}), every bound trajectory is
closed. Consequently, in the reduced 3 degrees of freedom
Hamiltonian~(\ref{eq:three}), every bound trajectory is also
closed. Evidently, the proof can be readily generalised to $N$ degrees
of freedom.

\subsection{A Sampler of Orbits}
\label{sec:numexpt}

It is interesting to investigate characteristic orbits corresponding
to the Hamiltonian ~(\ref{eq:three}).  Using a standard
Burlisch-Stoer code \cite{NR}, the motion of a unit mass particle was
followed in the octant with $x > 0$, $y > 0$ and $z > 0$. Stepsize and
tolerances were set to maintain accuracy to a level of around
$10^{-12}$ relative energy change, and integration lengths were
typically tens of periods.

\begin{figure}
\centering
\includegraphics[width=\textwidth]{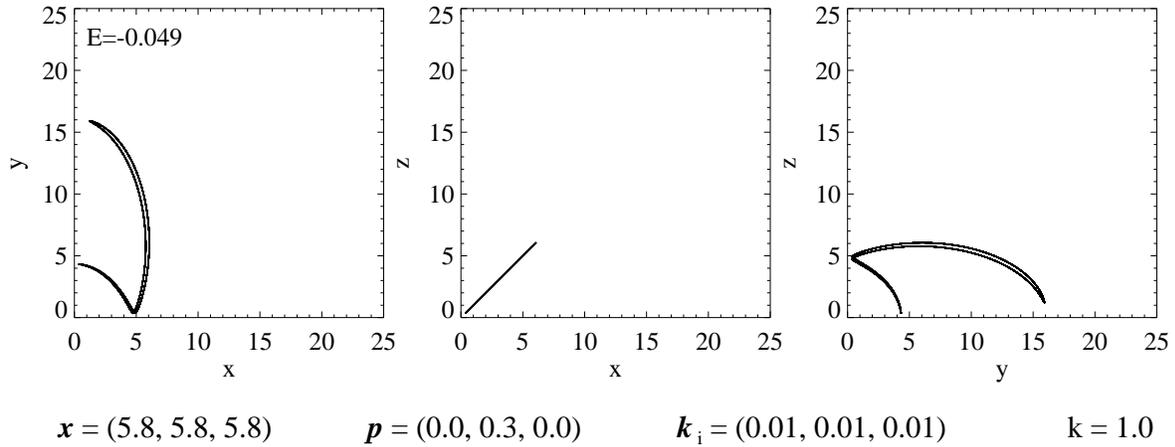}
\caption{
  A perturbed circular Keplerian orbit in the potential. The three
  panels show the projections in the $x-y$, $x-z$ and $y-z$ planes
  respectively from left to right. The initial conditions are given
  (to 2 s.f.)  below the plot and the energy in the top corner.  Note
  that although in this case the orbit lies in a plane this is not
  generally true.}
%since the centrifugal barriers have the same magnitude.}
\label{fig:orbit1}
\end{figure}

\begin{figure}
  \includegraphics[width=\textwidth]{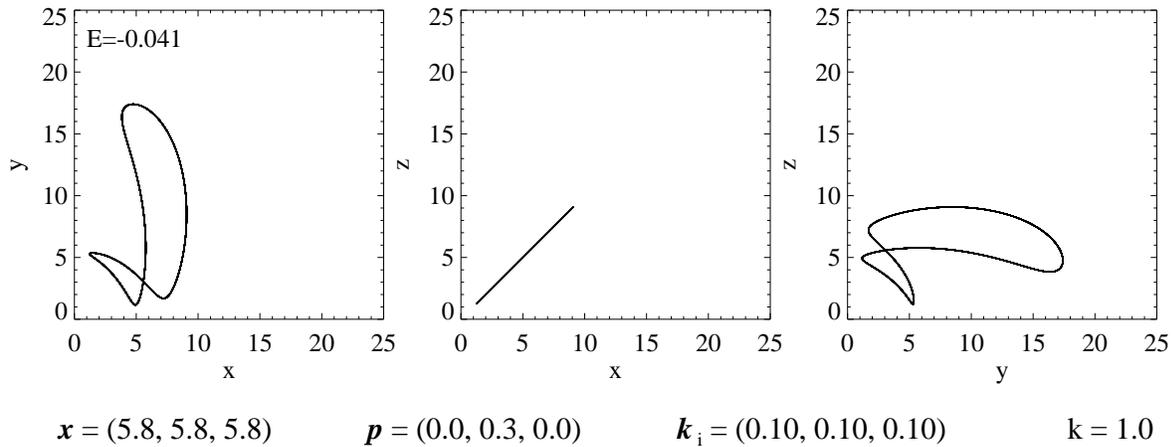}
\caption{
  As for Figure~\ref{fig:orbit1} but now the centrifugal barriers are
  an order of magnitude larger.}
\label{fig:orbit2}
\end{figure}

\begin{figure}
\includegraphics[width=\textwidth]{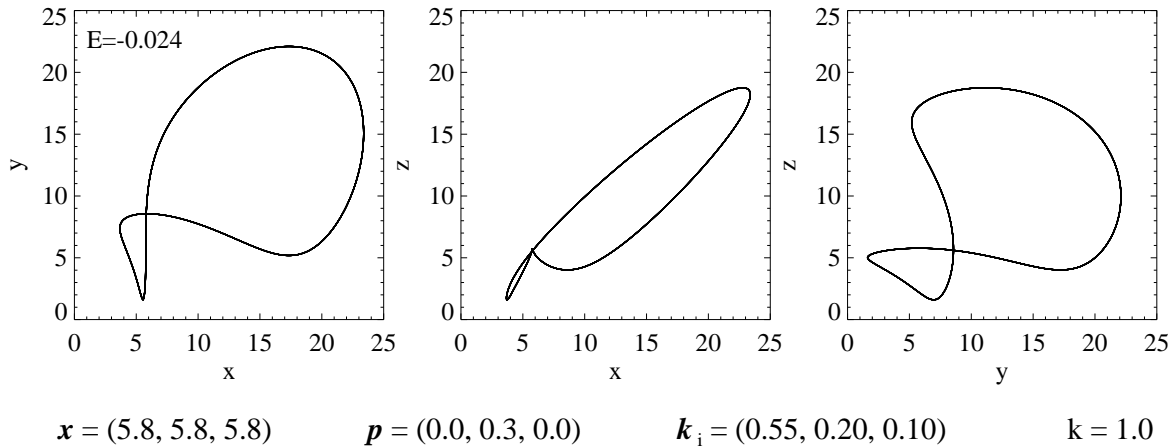}
\caption{
  As for Figure~\ref{fig:orbit2} but now the centrifugal barriers are
  different and the motion is no longer confined to a plane.}
\label{fig:orbit3}
\end{figure}

\begin{figure}
\includegraphics[width=\textwidth]{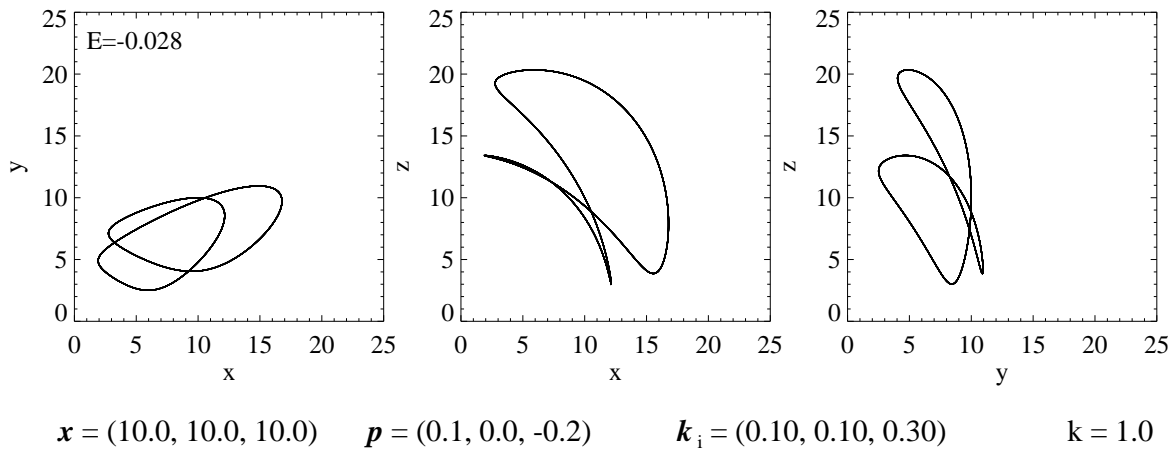}
\caption{
  This orbit in the potential has initial conditions that would place
  it on an elliptic orbit in the original Kepler problem.}
\label{fig:orbit4}
\end{figure}

Starting with initial conditions that would give a circular orbit in a
true Keplerian potential, the effect of relatively weak barriers can
be seen in Figure~\ref{fig:orbit1}. The orbit appears similar in shape
to the Keplerian ellipse but reflects off the three axes planes.  In
this case, as a consequence of the $k_i$ being equal and the symmetry
in the initial phase space position, the orbit is confined to a plane.
The effect of larger barriers can be seen through increasing the $k_i$
by a factor of ten, as shown in Figure~\ref{fig:orbit2}. The orbit is
now further distorted, and takes the form of a figure-of-eight.  If
the $k_i$ are not equal, the orbit is still closed, as shown in
Figures~\ref{fig:orbit3} and ~\ref{fig:orbit4}. The latter of these
two cases has a different set of initial conditions, which would place
it on an initially elliptical orbit in the true Keplerian problem.

Many more initial conditions and combinations of parameter values were
investigated. In all bound cases, every orbit is closed and strictly
periodic, which is reassuring confirmation of the existence of a fifth
isolating integral.

%%%%%%%%%%%%%%%%%%%%%%%%%%%%%%%FIGURE%%%%%%%%%%%%%%%%%%%%%%%%%%%%%%%%%
%%%%%%%%%%%%%%%%%%%%%%%%%%%%%%%%%%%%%%%%%%%%%%%%%%%%%%%%%%%%%%%%%%%%%%
%%%%%%%%%%%%%%%%%%%%%%%%%%%%%%%%%%%%%%%%%%%%%%%%%%%%%%%%%%%%%%%%%%%%%%
%%%%%%%%%%%%%%%%%%%%%%%%%%%%%%%%%%%%%%%%%%%%%%%%%%%%%%%%%%%%%%%%%%%%%%
%%%%%%%%%%%%%%%%%%%%%%%%%%%%%%%%%%%%%%%%%%%%%%%%%%%%%%%%%%%%%%%%%%%%%%

%%%%%%%%%%%%%%%%%%%%%%%%%%%%%%%%%%%%%%%%%%%%%%%%%%%%%%%%%%%%%%%%%%%%%%
%%%%%%%%%%%%%%%%%%%%%%%%%%%%%%%SECTION%%%%%%%%%%%%%%%%%%%%%%%%%%%%%%%%
%%%%%%%%%%%%%%%%%%%%%%%%%%%%%%%%%%%%%%%%%%%%%%%%%%%%%%%%%%%%%%%%%%%%%%

\section{The Integrals of Motion}
\label{sec:int}

Makarov et al.~\cite{Ma67} and Evans~\citep{Ev90} already showed that
all Hamiltonians of the form
\begin{equation}
H = \half|\u{\rdot}|^2 + F(r) + \frac{k_1}{x^2} + \frac{k_2}{y^2} 
+ \frac{k_3}{z^2}.
\label{eq:Vgensph}
\end{equation}
possess four isolating integrals of motion, arising from separability
of the Hamilton-Jacobi equation in the spherical polar and conical
coordinate systems. Here, $F(r)$ is an arbitrary function of the
spherical polar radius. The four isolating integrals are the energy
$E$ and three generalizations of the angular momentum components,
namely
\begin{eqnarray}
\label{eq:EtoI3}
E   &=& \frac{1}{2} |\mathbf{p}|^2 + F(r)
        + \frac{k_1}{x^2} + \frac{k_2}{y^2} + \frac{k_3}{z^2}    \\
I_1 &=& \frac{1}{2} L_1^2 + \frac{k_2 z^2}{y^2} 
         + \frac{k_3 y^2}{z^2} \\
I_2 &=& \frac{1}{2} L_2^2 + \frac{k_1 z^2}{x^2} 
         + \frac{k_3 x^2}{z^2} \\
I_3 &=& \frac{1}{2} L_3^2 + \frac{k_1 y^2}{x^2} 
         + \frac{k_2 x^2}{y^2}
\end{eqnarray}
where $\mathbf{p}$ and $\mathbf{L}$ are the linear momentum and
angular momentum vectors.  As Eq~(\ref{eq:Vgenkep}) is of this form,
four of the integrals of motion are already known and arise from
separability. The puzzle is that there is a fifth integral whose form
is unknown and which does not arise from separability.

In the case where one of the barriers, say that in the $x=y=0$ plane,
vanishes, the fifth integral is known to be \cite{Ev90}
\begin{equation}
\label{eq:I4_k30}
I_4 = L_1 p_2 - p_1 L_2 
       - 2 z \left(-\frac{k}{2 r} + \frac{k_1}{x^2} 
                   + \frac{k_2}{y^2} \right)
\end{equation}
and follows from separability in the rotational parabolic coordinate
system. If the fifth integral for the general problem
({\ref{eq:Vgenkep}) is quartic, it must reduce to the above integral
in the limit $k_3\to 0$. This suggests taking the ansatz
\begin{equation}
I_4 = \left(L_1 p_2 - p_1 L_2 - 2 z \left(-\frac{k}{2 r} +
\frac{k_1}{x^2} + \frac{k_2}{y^2} \right) \right)^2 + k_3
g(\mathbf{x},\mathbf{p})
\end{equation}
where $g(\mathbf{x},\mathbf{p})$ is a function of both position and
momentum yet to be determined.  Requiring the Poisson bracket of $I_4$
with the Hamiltonian to vanish leads to a solution for $g$. So, we
arrive at an isolating integral of the form
\begin{equation}
\label{eq:I4}
I_4 = \left( (\mathbf{L} \times \mathbf{p} ) _3 
             - 2 z \left (-\frac{k}{2 r} + \frac{k_1}{x^2} 
                          +\frac{k_2}{y^2} 
                          + \frac{k_3}{z^2} \right) \right)^2
      + \frac{2 k_3}{z^2} \left(\mathbf{r}\cdot\mathbf{p}\right)^2
\end{equation}
It is not yet proven that this is a functionally independent integral,
as it is possible to construct an infinite number of quartic integrals
from combinations of the four existing quadratic integrals. To test
for functional independence, the $5 \times 6$ Jacobian
\begin{equation}
\frac{\partial (E,I_1,I_2,I_3,I_4)}{\partial (x_i,p_i)}
\end{equation}
can be constructed and shown to be of rank 5. Thus, the integral given
in Eq~(\ref{eq:I4}) is the fifth functionally independent isolating
integral of motion for the Hamiltonian.

In the case when the $k_i$ are all zero, this integral reduces to the
$z$-component of the Laplace-Runge-Lenz vector.  In fact, if we work
through the same derivation but make the cyclic permutations $x \to y
\to z$, two more integrals are obtained, They are the equivalents of
Eq~(\ref{eq:I4}) with the coordinates permuted, and reduce to the $x-$
and $y-$components of the Laplace-Runge-Lenz vector. As is expected
they are not functionally independent, and this is easily demonstrated
by including them in the Jacobian and noting that it remains of rank
5. The three components are related through
\begin{equation}
I_{4_x} + I_{4_y} + I_{4_z} = 4 E (I_1+I_2+I_3+k_1+k_2+k_3)+k^2
\end{equation}
where $I_{4_z}$ is the integral given by Eq~(\ref{eq:I4}) and
$I_{4_x}$ and $I_{4_y}$ the cyclicly permuted versions.

%%%%%%%%%%%%%%%%%%%%%%%%%%%%%%%%%%%%%%%%%%%%%%%%%%%%%%%%%%%%%%%%%%%%%%
%%%%%%%%%%%%%%%%%%%%%%%%%%%%%%%SECTION%%%%%%%%%%%%%%%%%%%%%%%%%%%%%%%%
%%%%%%%%%%%%%%%%%%%%%%%%%%%%%%%%%%%%%%%%%%%%%%%%%%%%%%%%%%%%%%%%%%%%%%

\section{Action-Angle Variables}
\label{sec:acang}

It is also possible to solve this general problem in action-angle
variables. This is worthwhile as it provides insight into the
relationship our of our new superintegrable Hamiltonian with the
Kepler problem. Following \cite{Go80}, the actions can be shown to be
\begin{eqnarray}
J_\phi &=& \oint p_\phi d\phi = 2 \sqrt 2 \pi \left( \sqrt{I_2} - \sqrt{k_1}
- \sqrt{k_2}\right) \nonumber\\ 
J_\theta &=& \oint p_\theta d\theta = 2\pi \left( \sqrt{2I_1} - \sqrt{2
  I_2} - \sqrt{2k_3}\right) \\ 
J_r &=& \oint p_r dr = 2\pi \left(-\sqrt{2I_1} -
\frac{k}{\sqrt{-2 E}} \right)\nonumber
\label{eq:acts}
\end{eqnarray}
and hence
\begin{equation}
E = \frac{-2k^2 \pi^2}{J_r+J_\theta + J_\phi 
           + 2\sqrt{2}\pi(\sqrt{k_1}+\sqrt{k_2}+\sqrt{k_3})^2}
\end{equation}
As expected, the Hamiltonian depends on the actions only through the
combination $J_r + J_\theta + J_\phi$, implying that the three
frequencies of the classical motion are the same. If ($w_r, w_\theta,
w_\phi$) are the angles conjugate to~(\ref{eq:acts}), then we can make
a canonical transformation to new action-angle coordinates ($J_1,
J_2,J_3, w_1, w_2, w_3$), using the generating function
\begin{equation}
F = (w_\phi-w_\theta)J_1 + (w_\theta - w_r)J_2 + w_r J_3
\end{equation}
The new actions are related to the old via
\begin{equation}
J_1 = J_\phi, \qquad J_2 = J_\theta + J_\phi, \qquad J_3 = J_r +
J_\theta + J_\phi,
\end{equation}
and so the Hamiltonian becomes
\begin{equation}
H = E = \frac{-2k^2 \pi^2}{J_3 + 2\sqrt{2}\pi(\sqrt{k_1}
                                               +\sqrt{k_2}+\sqrt{k_3})^2}
\end{equation}
and depends on only one of the new actions. Using Hamilton's
equations, we see that the angle $w_3$ increases linearly with time,
whilst the angles $w_1$ and $w_2$ are the additional integrals of
motion. They can be found by explicit construction of Hamilton's
characteristic function $S$
\begin{equation}
S = \int p_\phi d\phi + \int p_\theta d\theta + \int p_r dr
\end{equation}
followed by use of the equations $w_i = \partial S / \partial J_i$
(see \cite{Go80} for the equivalent calculation for the Keplerian
potential).  We find that
\begin{eqnarray}
w_1 &=& \frac{1}{4\pi} \arccos \left(\frac{A^2 \cos 2\phi - k_1 + k_2}
                                    {\sqrt{(A^2-k_1+k_2)^2-4A^2k_2}}\right) 
\nonumber \\
{}& &   - \frac{1}{4\pi} \arccos 
                             \left(\frac{2A^2\cot^2\theta - B^2 + A^2 +k_2}
                             {\sqrt{(B^2-A^2-k_2)^2 - 4A^2k_2}}
                                           \right) \\
w_2 &=& \frac{1}{4\pi} \arcsin 
                             \left(\frac{B^2 \cos 2\theta + A^2 - k_3}
                                         {\sqrt{(A^2+B^2-k_3)^2-4A^2B^2}}
                                                    \right) \nonumber \\
{}& &  - \frac{1}{2\pi} \arcsin \left(\frac{k r - 2 B^2} 
                                              {r\sqrt{k^2 +4 B^2 E}}\right) 
\end{eqnarray}
where
\begin{eqnarray}
A &=& \frac{J_1}{2\sqrt{2}\pi} +\sqrt{k_1} +\sqrt{k_2} = \sqrt{I_2} \\
B &=& \frac{J_2}{2\sqrt{2}\pi} +\sqrt{k_1} +\sqrt{k_2}+\sqrt{k_3} = \sqrt{I_1}
\end{eqnarray}
Note that in the case that $k_1 = k_2 = k_3 =0$ (the Kepler problem),
it is usual to introduce the inclination $i = \arccos (A/B)$ of the
orbital plane. Tnen, the angle $w_1$ reduces to
\begin{equation}
w_1 =  \frac{1}{2\pi} \left(\phi - \arcsin \left(\cot \theta \cot
i\right)\right)
\end{equation}
which is the longitude of the ascending node. The second angle $w_2$
is easiest evaluated in the orbital plane with polar coordinates ($r,
\varphi$) and becomes
\begin{equation}
w_2 = \frac{1}{2\pi} \left( (\varphi - \varphi_{\rm lan}) -
(\varphi - \varphi_{peri}) \right) = \omega.
\end{equation}
The first integral is therefore the angular difference between the
orbital position and the longitude of the ascending node, the second
the angular difference between the orbital position and the periapse.
Thus, $w_2$ reduces to $\omega$, the longitude of the periapse in the
Kepler problem.

%%%%%%%%%%%%%%%%%%%%%%%%%%%%%%%%%%%%%%%%%%%%%%%%%%%%%%%%%%%%%%%%%%%%%%
%%%%%%%%%%%%%%%%%%%%%%%%%%%%%%%SECTION%%%%%%%%%%%%%%%%%%%%%%%%%%%%%%%%
%%%%%%%%%%%%%%%%%%%%%%%%%%%%%%%%%%%%%%%%%%%%%%%%%%%%%%%%%%%%%%%%%%%%%%

\section{Summary and Conclusions}
\label{sec:con}

We have found a new superintegrable Hamiltonian, which is a
generalization of the well-known Kepler problem. There are five
isolating integrals of the motion, namely the energy and
generalizations of the componenets of the angular momentum and
Laplace-Runge-Lenz vectors. Intriguingly, some of the integrals of
motion are quartic in the momenta and do not arise from separability
of the Hamilton-Jacobi equation.

There are three interesting questions which merit further
research. First, it is clear that the $N$ degrees of freedom
Hamiltonian
\begin{equation}
H = \half \sum_{i=1}^N p_i^2 - {k\over r} + {k_i \over x_i^2}
\label{eq:con}
\end{equation}
is also superintegrable. It would be interesting to find the complete
set of $2N-1$ functionally independent integrals of motion in this
case.  Second, it is well-known \cite{Fo35,Ba36} that the additional
integrals of motion in the Kepler problem arise from the existence of
the dynamical symmetry group SO(4). It would be interesting to
understand the group theoretic interpretation of the integrals of
motion discussed in this paper. Third, although the reduction
technique we used to generate the superintegrable potential in Section
\ref{sec:evi} is simple, it can be made to do some more work. For
example, it is also clear that the Hamiltonian [c.f., eq~(\ref{eq:anosc})]
$$H = \half (p_1^2 + p_2^2 +p_3^2) + \ell^2 x^2 + m^2 y^2 + n^2 z^2
+ {k_1 \over x^2} + {k_2\over y^2} + {k_3 \over z^2} .$$
always has 5 independent integrals of motion as well, whose form
remains to be established. Perhaps all superintegrable potentials in
three degrees of freedom can be viewed as projections of higher
dimensional Keplerian or harmonic oscillator motion?

%%%%%%%%%%%%%%%%%%%%%%%%%%%%%%%%%%%%%%%%%%%%%%%%%%%%%%%%%%%%%%%%%%%%%%
%%%%%%%%%%%%%%%%%%%%%%%%%%%%BIBLIOGRAPHY%%%%%%%%%%%%%%%%%%%%%%%%%%%%%%
%%%%%%%%%%%%%%%%%%%%%%%%%%%%%%%%%%%%%%%%%%%%%%%%%%%%%%%%%%%%%%%%%%%%%%


\begin{thebibliography}{00}

\bibitem[Landau \& Lifshitz(1969)]{La69} Landau, L., \& Lifshitz,
  E.M., 1968, Mechanics, Pergammon Press, Oxford, p. 151ff

\bibitem[Sommerfeld(1923)]{So23} Sommerfeld, A., 1923, Atomic
  Structure and Spectral Lines, Methuen, London, p. 118ff

\bibitem[Born(1927)]{Bo27} Born, M., 1927, Mechanics of the
Atom, G. Bell, London, p. 265ff

\bibitem[Fris et al.(1965)]{Fr65} Fris, J., Mandrosov, V., 
Smorodinsky, Y.~A., Uhl{\'{\i}}, M., \& Winternitz, P.\ 1965, Physics 
Letters, 16, 35

\bibitem[Makarov et al.(1967)]{Ma67} Makarov, A. A., Smorodinsky,
  Y.~A., Valiev K., \& Winternitz, P., \ 1967 Nuovo Cimento 52, 1061.

\bibitem[Evans(1990)]{Ev90} Evans, N.~W.\ 1990, Phys. Rev. A, 41, 5666

\bibitem[Tempesta et al.(2005)]{Te05} Tempesta, P., Winternitz, P.,
Harnad, J., Miller Jr, W., Pogosyan, G., Rodriguez, M., 2005,
Superintegrability in Classical and Quantum Systems, American
Mathematical Society

\bibitem[Gravel \& Winternitz(2002)]{Gr02} Gravel, S., Winternitz,
  P. 2002, J Math Phys, 43, 5902

\bibitem[Jauch \& Hill(1940)]{Ja40} Jauch, J.M., \& Hill, E.L., 1940,
  Phys. Rev., 57, 641

\bibitem[Adler(1977)]{Ad77} Adler, M., 1977, Comm. Math. Phys., 55, 195

\bibitem[Wojciechowski(1983)]{Wo83} Wojciechowski, S.\ 1983, Physics
   Letters A, 95, 279

\bibitem[Moser(1970)]{Mo} Moser, J., 1970, Comm. Pure and Applied
  Maths, 23, 609

\bibitem[Press et al.(2002)]{NR}
Press, W.~H., Teukolsky, S.~A., Vetterling, W.~T., \& Flannery, B.~P.\ 2002,
Numerical recipes in C++ : the art of scientific computing, 
Cambridge University Press

\bibitem[Goldstein(1980)]{Go80} Goldstein, H., \ 1980, Classical
  Mechanics, Addison-Wesley, Reading, p. 613-615

\bibitem[Fock(1935)]{Fo35} Fock, V., \ 1935, Z. Phys., 98, 145

\bibitem[Bargmann(1936)]{Ba36} Bargmann, V., \ 1935, Z. Phys., 99, 576


\end{thebibliography}
\end{document}